\documentclass{ws-mpla}
\usepackage{amsfonts}
\usepackage{amsbsy}
\usepackage{mathrsfs}
\usepackage{graphicx}

\begin{document}

\catchline{}{}{}{}{}

\title{Estimate of the magnetic field strength in heavy-ion collisions}

\author{\footnotesize V.~V.~SKOKOV}
\address{Gesellschaft
 f\"ur Schwerionenforschung mbH, Planckstr. 1,
D-64291 Darmstadt, Germany\\
Frankfurt Institute for
Advanced Studies, Universit\"at Frankfurt, D-60438 Frankfurt am
Main, Germany\\
Joint Institute for Nuclear Research,
RU-141980 Dubna, Moscow Region, Russia
}

\author{A.~Yu.~ILLARIONOV}
\address{Dipartimento di Fisica dell'Universit\'{a} di Trento,
via Sommarive 14, I--38050 Povo, Trento, Italy}

\author{V.~D.~TONEEV }
\address{Gesellschaft
 f\"ur Schwerionenforschung mbH, Planckstr. 1,
D-64291 Darmstadt, Germany\\
Joint Institute for Nuclear Research,
RU-141980 Dubna, Moscow Region, Russia}

\maketitle

\pub{Received (Day Month Year)}{Revised (Day Month Year)}

\begin{abstract}
Magnetic fields created in the noncentral heavy-ion collision are studied
within a microscopic transport model, 
namely the Ultrarelativistic Quantum
Molecular Dynamics model (UrQMD). Simulations were carried out for different
impact parameters within the SPS energy range  ($E_{lab} = 10 - 158 A$ GeV)
and for highest energies accessible for RHIC.  We show that the magnetic field
emerging in heavy-ion collisions has the magnitude of the order of
$eB_y \sim10^{-1} \cdot  m_\pi^2$ for the SPS energy range and $eB_y \sim m_\pi^2$
for the RHIC energies. The estimated value of the  magnetic field strength for
the LHC energy amounts to  $eB_y\sim 15 \cdot  m_\pi^2$.
\end{abstract}

\date{\today}
\ccode{PACS Nos.: 25.75.-q, 25.75.Ag}

\section{Introduction}
Heavy-ion collisions are intensively investigated both experimentally and
theoretically to reveal information about properties of the nuclear
matter under extreme conditions.  Heavy-ion collision experiments at
AGS (BNL), SPS (CERN) and RHIC (BNL) as well as future ones
 planned at FAIR (GSI), NICA (JINR), LHC (CERN) make it possible to
explore the phase diagram  of the nuclear matter in a broad parameter 
range of temperature and baryon density. 
One of the most important
issue addressed in nucleus-nucleus collisions is the possibility
for nuclear matter to undergo phase transitions into a new state
of matter. At least two of these transitions, chiral and deconfinement,
are commonly expected. Though at zero baryon chemical potential both
transitions seem to have the same critical temperature, it is not clear at the
moment whether this situation will remain at the large chemical potential or the the
chiral phase transition anticipates the deconfinement one, allowing for the 
``quarkyonic'' states~\cite{McLerran:2007qj,Hidaka:2008yy}
with the massless confined quarks.
Despite the recent progress~\cite{Bazavov:2009zn,Aoki:2009sc}, 
the lattice QCD calculations at finite baryon chemical potential
do not provide fully reliable information on the subject.

One of the most exciting signals of the deconfinement and
the chiral phase transitions in heavy-ion collisions,
the chiral magnetic effect, suggested in \cite{Kharzeev:2007jp},
predicts the preferential emission of
charged particles along the direction of angular momentum in the case of 
the noncentral heavy-ion collisions due to the presence of nonzero chirality.
As it was stressed in~\cite{Kharzeev:2007jp,Fukushima:2008xe},
both the deconfinement and chiral phase transitions 
are the essential requirements 
for the chiral magnetic effect to take place. The first one is needed, since
only in this case the soft quarks can be separated
by a distance larger than the radius of a nucleon. The second one is also
required~\cite{Kharzeev:2007jp,Fukushima:2008xe}, since nonzero values
of the chiral condensate drive asymmetry between the number
of right- and left-handed quarks to zero.

The effects caused by a strong magnetic field are not limited by
the chiral magnetic effect \cite{Fukushima:2008xe}. They include also the
induced chiral symmetry breaking~\cite{Gusynin:1995gt},
modification of the nature of the chiral phase transition (e.g. turning
the crossover phase transition to the first-order one through influence on the chiral
condensate~\cite{Cohen:2007bt,Buividovich:2008wf,Fraga:2008qn,Mizher:2008dj,Agasian:2008tb,Ayala:2009ji}),
influence on the possible color-conducting 
phases~\cite{Alford:1999pb,Fukushima:2007fc,Noronha:2007wg} 
and the pion condensate~\cite{Voskresensky:1980nk},
spontaneous creation of the axial currents~\cite{Son:2004tq,Metlitski:2005pr},
and formation of the $\pi_0$-domain walls~\cite{Son:2007ny}. Recently, the effect of
a large magnetic field on the sound velocity of a propagating plane wave was
studied in Ref.~\cite{Sadooghi:2009fi}.

The key quantity of these  effects is a magnitude of the background magnetic field
strength created in heavy-ion collisions. 
The early estimate of the magnetic
field for the RHIC energy was made in Ref.~\cite{Kharzeev:2007jp}. It was shown
that the field may reach very high values
$eB\sim3 \cdot m_\pi^2 \sim 3 \times 10^{18}$ Gauss\footnote{Probably the first rough
estimate of the magnetic field in heavy-ion collision was made in Ref.~\cite{Voskresensky:1980nk}}.
The aim of this article is to improve a qualitative estimate, make the
quantitative calculation  of the magnetic field for heavy-ion collisions 
at different impact parameters and different energies, study its 
characteristics. The calculations will be carried out
within the microscopic transport model, namely the Ultrarelativistic Quantum Molecular
Dynamics model (UrQMD), and will be complemented by analytical considerations.

\begin{figure}[t]
\centerline{\includegraphics[height=3.0truecm] {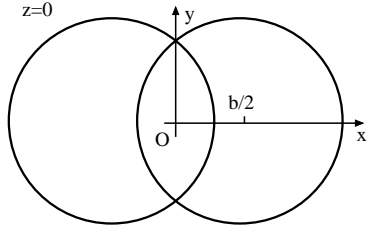}}
\caption{ The transverse plane of a noncentral heavy-ion collision.
The impact parameter of the collision is $b$. The magnetic field
is to be calculated at the origin $O$ and along the $y$-axis.
}
\label{tr}
\end{figure}

\section{Magnetic field in heavy-ion collisions}

The Ultrarelativistic Quantum Molecular Dynamics model
is a microscopic model used to simulate (ultra)relativistic heavy-ion collisons
in the energy range from Bevalac and SIS up to  AGS, SPS and RHIC. The detailed
description of the model can be found in Refs.~\cite{Bass:1998ca,Bleicher:1999xi}.

In this paper we focus our investigation on the magnetic field evaluated
in the center of the created fireball and on the energy density of nuclear matter
in the region surrounding this point.

The magnetic field strength at a position $\vec{x}$ and time $t$ is defined by
the Lienard-Wiechert potentials
\begin{equation}
e \vec{B}(t, \vec{x}) = \alpha_{\rm{EM}} \sum_n Z_n  \frac{1-v_n^2}{\left(R_n - \vec{R}_n\vec{v}_n\right)^{3}} \left[ \vec{v}_n \times \vec{R}_n  \right],
\label{Linard}
\end{equation}
where the fine-structure constant $\alpha_{\rm{EM}}\approx1/137$,
$Z_n$ is an electric charge of the $n$th particle (in units of the electron
charge) and $\vec{R_n} = \vec{x}-\vec{x}_n$ is a radius vector of particle,
$\vec{v}_n$ is a particle velocity.
The quantities $\vec{v}_n$ and $\vec{x}_n$ are taken at the time moment
$t'$ retarded with respect to the observation time $t$ to be defined implicitly
by the following equation
\begin{equation}
| \vec{x}-\vec{x}_n(t') | + t' = t~,
\label{retarded}
\end{equation}
where  we use the natural units and set $c=1$.
Summation is to be carried out over all charged particles.
However, to avoid uncertainties coming from participant contribution,
the summation was performed only over the spectators in our calculations.

In principle, the  magnetic field can be calculated for any  space points.
We limit ourselves only to evaluation of the field
at the central point of the reaction volume $O$
(see Fig.~\ref{tr}) and the dependence on $y$-coordinate at $z=x=0$.

On the basis of the expression (\ref{Linard}) let us draw several conclusions for
the main properties of the magnetic field at the origin $O$. First of all, it is
evident from symmetry reasons that the magnetic field will be negligible  for
a collision with a small impact parameter. From the same symmetry
considerations one obtains that the field will have only nonzero $B_y$-component.
Second, the field will be negligible for low bombarding energies because the 
field strength is proportional to the particle velocity. On the other hand,
for very high ultrarelativistic energies of a collision  the contribution 
to the magnetic field is feasible only for particles
close to the transverse plane $\left(R_n - \vec{R}_n\vec{v}_n\right)\sim0$.
The contribution from particles away from the transverse plane is suppressed
by the factor $(1-v^2)$.
From the expression (\ref{Linard}) it also follows that the characteristic magnetic
field dependence on nuclei charge is given by $eB\sim Z/R^2$, where $R$ is
the characteristic length scale proportional to the nuclei radius.
For stable nuclei we have $R\sim A^{1/3}\sim Z^{1/3}$ and thus $eB\sim Z^{1/3}$,
demonstrating a weak field dependence on the nuclei charge (see similar estimate in
Ref.~\cite{Voskresensky:1980nk}).

\begin{figure}
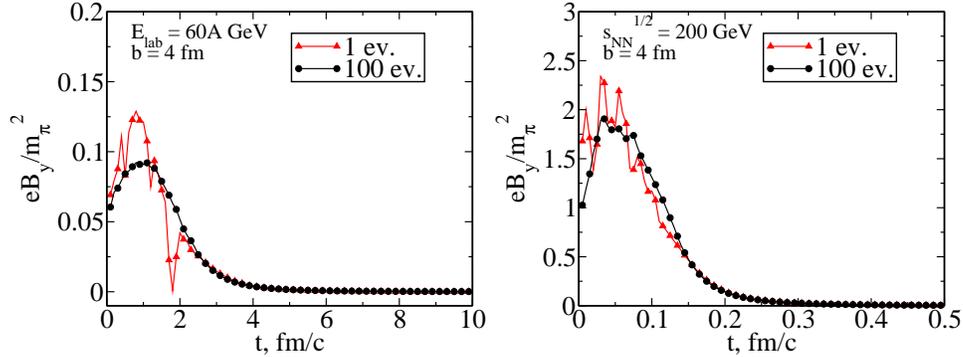

\centerline{
\includegraphics[height=4.7truecm] {Elab_60_4_3c.eps}
\includegraphics[height=4.7truecm] {s_200_4_3c.eps}}
\caption{The time evolution of
the magnetic field strength $eB_y$ at the central point $O$ (see Fig.\ref{tr})
in  Au-Au collisions
with impact parameter $b=4$ fm in the UrQMD model, in one event (``1 ev.'') and
averaged over 100 events (``100 ev.'').
The symbols are plotted every $\Delta t=0.2$ fm/c for $E_{lab} = 60A$ GeV and
$\Delta t=0.01$ fm/c  for $\sqrt{s_{NN}}=200$ GeV.
}
\label{B_1_100}
\end{figure}

\begin{figure}
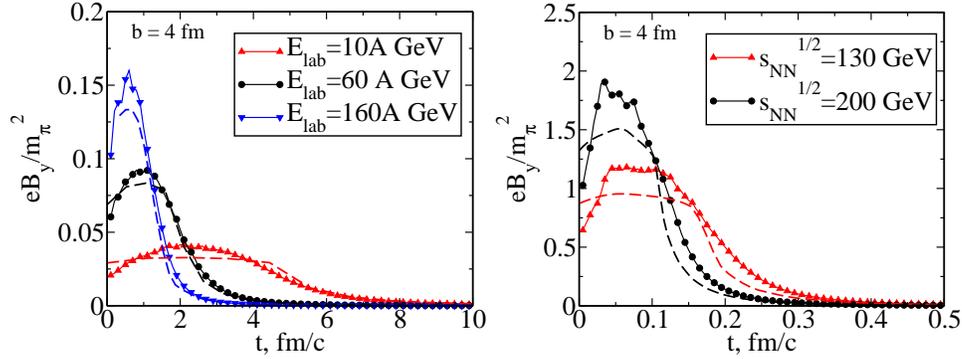

\centerline{
\includegraphics[height=4.7truecm] {urqmd_10_60_160.eps}
\includegraphics[height=4.7truecm] {urqmd_130_200.eps}}
\caption{The time evolution of
the magnetic field strength $eB_y$ at the central point $O$ (see Fig.\ref{tr})
in  Au--Au collisions  with impact parameter, $b=4$ fm, in the UrQMD model,
for different  bombarding energies. The symbols are plotted every
$\Delta t=0.2$ fm/c for $E_{lab} = 60A$ GeV and $\Delta t=0.01$ fm/c
for $\sqrt{s_{NN}}=200$ GeV. The magnetic field obtained by modelling the gold ions
as two Lorenz contracted non-interacting
uniformly charged spheres with radius $R=7$ fm are  shown by dashed lines.
}
\label{B_SPS_an}
\end{figure}

In Figs.~\ref{B_1_100} and \ref{B_SPS_an}, the time evolution of
the magnetic field  strength for SPS and RHIC energies is shown.
The magnetic field is created in the 
noncentral Au--Au collision with the impact
parameter $b=4$ fm. The resulting field strength is averaged over 100 events
to reduce statistical fluctuations. It is clear, however, that the magnetic
field in one individual event can be significant for an observable effect.
The nonzero components of the magnetic field strength $eB_y$ in a single event 
are plotted on the same figures to demonstrate its deviations from the average
 value, see Fig.~\ref{B_1_100}.
In Fig.~\ref{B_SPS_an}, the magnetic field strength $eB_y$
was estimated also assuming a collision of two uniformly charged Lorenz
contracted noninteracting spheres with the radius $R=7$ fm. The spheres move with
the velocity defined by the collision energy. The results obtained in this
semianalytical model and the UrQMD one agree to each other. The magnitude of the
magnetic field estimated in Ref.~\cite{Kharzeev:2007jp} for an earlier stage of
Au--Au collision at the RHIC energy $\sqrt{s_{NN}}=200$ GeV and the impact parameter
$b=4$ fm is about $eB \approx 1.3\cdot m_\pi^2$, which is close
to our calculations\footnote{We measure the magnetic field strength in the units
of the pion mass squared, using for definiteness $m_\pi=140$ MeV.
However, the magnetic field strength can be translated into the CGS system by
the following identity
$
m_\pi^2  =  140^2 \times 0.512 \cdot 10^{14} {\rm Gauss}  \approx 10^{18}   
{\rm Gauss}. $
}.  Note that this magnetic field strength is higher by about 4 orders of magnitude
than that in the surface of magnetar~\cite{magnetar}. It is impossible to make steady
fields stronger than $4.5\cdot 10^5$ Gauss in the lab because the magnetic stresses
of such fields exceed the tensile strength of terrestrial materials.

\begin{figure}
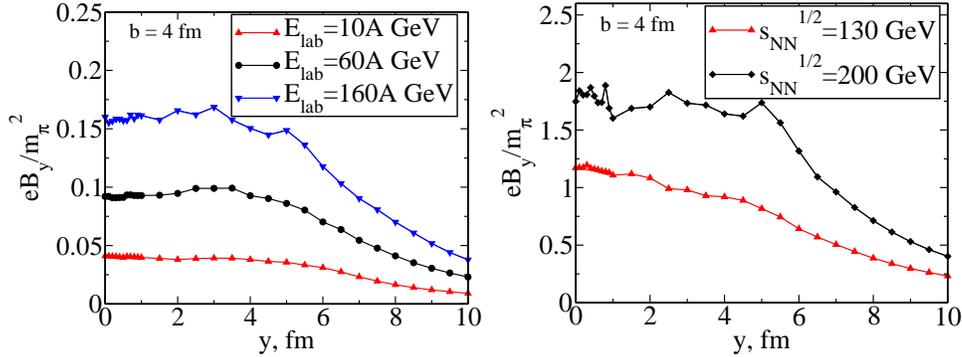

\centerline{
\includegraphics[height=4.7truecm] {urqmd_Y.eps}
\includegraphics[height=4.7truecm] {urqmd_Y_RHIC.eps}}
\caption{
The dependence of the magnetic field on the coordinate $y$.
The calculation was carried out for Au--Au collisions
with impact parameter $b=4$ fm in the UrQMD model.
The magnetic field  is taken at fixed time corresponding to its maximum
value at $y=0$.
}
\label{Y}
\end{figure}

In Fig.~\ref{Y} we show the dependence of the magnetic field strength on the
coordinate $y$ for a fixed time corresponding to its maximal value at $y=0$.
As it is seen, the field stays approximately constant up to
$y\sim5$ fm, demonstrating the high degree of homogeneity 
in the central region.

\begin{figure}
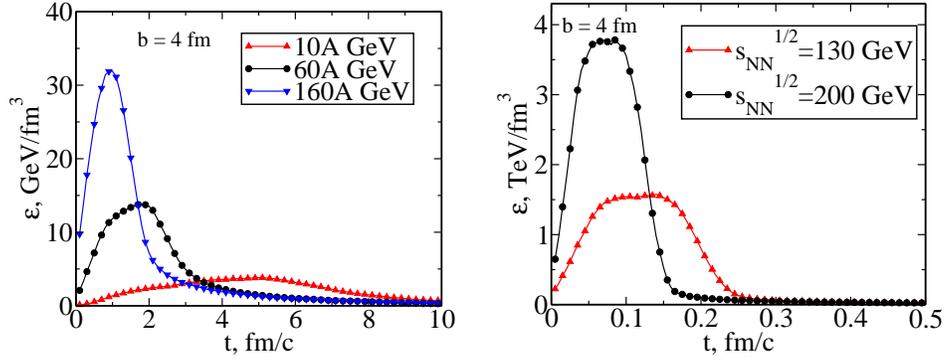

\centerline{
\includegraphics[height=4.7truecm] {urqmd_Box_10AGeV_4.eps}
\hspace{0.3cm}
\includegraphics[height=4.7truecm] {urqmd_Box_RHIC.eps}}
\caption{
The time evolution of the energy density in the central region (see text for 
details) for SPS and RHIC energies.
}
\label{epsB}
\end{figure}

The large value of the magnetic field strength itself does not guarantee
possible observable effects, but additional requirements are needed.
For the chiral magnetic effect the system should be in the deconfinement
and the chiral restored phase. To demonstrate that matter in the central
region is presumably in the QGP phase, we calculate the energy density.
We define the center region of the created fireball
as a Lorentz contracted box with the transverse and longitudinal size
$l_x=l_y=4$ fm, $l_z=(4/\gamma_{cm})$ fm, where $\gamma_{cm}$ is a Lorentz
factor calculated in the center of mass frame. The energy density is defined as
 $\varepsilon=E/V_{box}$, where $V_{box} = 64/\gamma_{cm}$ fm$^3$
and $E$ is the energy of particles in the box
\begin{equation}
E = \sum_{(x,y,z)\in\rm{box}}  \sqrt{p_i^2+m_i^2} \,.
\label{energy}
\end{equation}
The calculated time evolution of the energy density in the central region is shown
in Fig.~\ref{epsB}. Even for the bombarding energy $E_{lab}=10A$ GeV and the impact
parameter $b=4$ fm, the energy density in the central region is sufficient to reach
the QGP phase. However, the question whether the system is in local equilibrium and
can be described by the temperature and chemical potential(s) is still open.
In the framework of the UrQMD model for central collisions the equilibration
was considered in Refs.~\cite{Bravina:1999dh,Bravina:2007zza}.

\begin{figure}
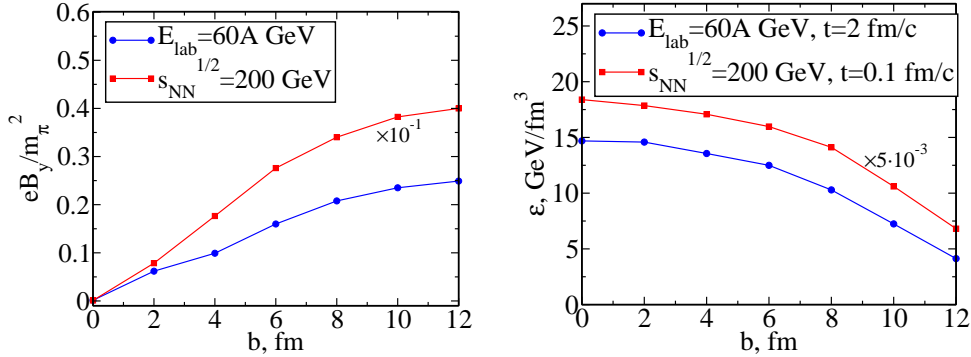

\centerline{
\includegraphics[height=4.7truecm] {urqmd_B.eps}
\hspace{0.3cm}
\includegraphics[height=4.7truecm] {urqmd_box.eps}}
\caption{
The magnetic field (left panel) and the energy density (right panel) dependence on 
the impact parameter $b$ for $E_{lab} = 60A$ GeV and $\sqrt{s_{NN}} = 200$ GeV.
}
\label{impact}
\end{figure}

The magnetic field and energy density dependence on the
impact parameter are depicted in Fig.~\ref{impact} for Au--Au collisions
at bombarding energies $E_{lab}=60A$ GeV and $\sqrt{s_{NN}}=200$ GeV . It is shown
that the smaller the impact parameter $b$ the larger the energy density $\varepsilon$ 
and the smaller the magnetic field $eB_y$ is, as it could be expected from  simple
symmetry considerations.

\section{Conclusions}
The magnitudes of the magnetic field 
strength play an important role in estimations
of possible observable effects of the deconfinement and chiral phase transitions
in heavy-ion collisions. The magnetic field strength in heavy-ion collisions
at RHIC energies was estimated by D. Kharzeev, L. McLerran and H. Warringa
following simple considerations, see~\cite{Kharzeev:2007jp} for details.
For energy $\sqrt{s_{NN}}=200$ GeV, the authors obtained the magnetic field
$eB_y\approx 3\cdot m_\pi^2$ which is strong enough to provide necessary
conditions for the effect to be measured~\cite{Voloshin}. In this paper
we employed the UrQMD model
for calculation of the magnetic field strength at various bombarding energies
$E_{lab}=10,60,160A$ GeV and $\sqrt{s_{NN}}=130, 200$ GeV. We showed the time evolution
of the magnetic field, its dependence on the transverse coordinate $y$ and
the impact parameter $b$. The obtained results demonstrate that the fields emerging
in heavy-ion collision has a high degree of homogeneity in the central region.
The analysis of the energy density in the central region allows us
to draw the conclusion that the considered system may reach the deconfinement and
chiral restored phase. To all appearance, the requirements to observe the
chiral magnetic effect are fully satisfied. We compared the UrQMD calculations
with a simple semianalytical model considering
heavy ions as two Lorenz-contracted uniformly-charged noninteracting spheres.
The semianalytical results are in a satisfactory agreement 
with the dynamical ones. The
former underestimate the UrQMD calculations by only few percent. Using this
model we can estimate the lowest bound of the maximal magnetic field strength
at the LHC energy  $\sqrt{s_{NN}}=4.5$ TeV to be about  $15\cdot  m_\pi^2$
in collisions of Pb--Pb ions with the impact parameter $b=4$ fm, which is twice
higher than that estimated in~Ref.~\cite{Fraga:2008qn}.

\section{Acknowledgements}
We are grateful to K.K. Gudima, E.E. Kolomeitsev and D.N.  Voskresensky
for criticism and valuable discussions.  We are thankful to D.E. Kharzeev and
O.V. Teryaev for reading the manuscript. V.S. and V.T.  were partially supported
by the Bundesministerium f\"ur Bildung und Forschung (BMBF project RUS 08/038).
V.S. acknowledges the support of the  RFBR grant No. 08-02-01003-a and V.T.
 acknowledges the Heisenberg-Landau grant support.

\end{document}